\documentclass[aps,onecolumn,prd,nofootinbib]{revtex4}
\usepackage{amsmath}
\usepackage{graphicx}
\usepackage{dcolumn}
\usepackage{bm}
\usepackage{amssymb}
\usepackage{latexsym}

\newcommand{\be}{\begin{equation}}
\newcommand{\ee}{\end{equation}}
\newcommand{\bq}{\begin{eqnarray}}
\newcommand{\eq}{\end{eqnarray}}

\begin{document}


\title{Stability of de Sitter solutions sourced by dark spinors}

\author{Guoying Chee}
\affiliation{Physics Department, Liaoning Normal University, Dalian
116029, China} \affiliation{Purple Mountain Observation, Academia
Sinica, Nanjing 210008, China}

\begin{abstract}
Cosmology of ELKO and Lorentz Invariant NSS has been investigated
using dynamical system method starting from the proposal of Boehmer
et al [arXiv:1003.3858]. Some different results have been obtained
by a different approach. The exact solutions described by fixed
points of the dynamical system and their stability have been
discussed. Some stable solutions corresponding to de Sitter
universes have been obtained.
\end{abstract}

\maketitle

\section{Introduction}

Recently a new class of spinors refer to as dark spinors or ELKO has been
found [1] and has received quite some attention [2]. Their effects in
cosmology have been investigated [3,4,5]. Their dynamics is more general
that than that of Dirac or Majorana spinors, even when self-interactions are
taken into account. This leads to a more general and thus more interesting
cosmological behavior than that exhibited by normal spinors, including for
instance the existence of non-trivial de Sitter solutions. However, ELKOs
spinors are not Lorentz invariant and their definition requires a preferred
space-like direction. In a recent article[6], a new class of theories of
non-standard spinors (NSS) has been constructed. By providing a general
treatment of class of NSS models based on an action principle a Lorentz
invariant, ghost-free, non-local spinor field theory has been developed. The
cosmological applications of both the original ELKO and the Lorentz
invariant NSS have been examined respectively. Especially, and the existence
of non-trivial de Sitter solutions in each case has been discussed.

The cosmological dynamics of the effective scalar degree of freedom in both
ELKO and Lorentz invariant NSS cosmology show a large number of very
interesting properties. The cosmological evolution of the NSS energy density
exhibits a much wider range of behavior than that seen with Dirac spinors.
The existence of stable de Sitter solutions means that NSS could represent
an alternative to scalar field inflation or dark energy.

In this paper we investigate the evolution of a model of the
Friedmann-Robertson-Walker (FRW) universe in both ELKO and Lorentz invariant
NSS cosmology using the dynamical systems approach [7, 8, 9]. Some results
different from [6] will be given.

\section{ELKO Cosmology}

Consider a setting in which we only have ELKO spinor fields coupled
minimally to gravity, this means we neglect all possible interaction terms.
We begin by focusing on the background of a flat FRW spacetime with line
element:

\[
ds^2=dt^2-a^2(t)d{\bf x}^2.
\]

A ELKO spinor can be defined by $\psi =\varphi (t)\xi $, with $\stackrel{%
\lnot }{\xi }\xi =1$, where $\xi $ is a constant spinor, and then $\varphi $
can be treated as the only dynamical variable cosmological. The energy
density $\rho _\varphi $, and pressure $p_\varphi $, are, respectively [6],
\begin{eqnarray}
\rho _\varphi &=&\left[ \frac 12\stackrel{\cdot }{\varphi }^2+V(\varphi
^2)\right] +\frac 38H^2\varphi ^2,  \nonumber \\
p_\varphi &=&\left[ \frac 12\stackrel{\cdot }{\varphi }^2-V(\varphi
^2)\right] -\frac 38H^2\varphi ^2-\frac 14\stackrel{\cdot }{H}\varphi ^2-%
\frac 12H\varphi \stackrel{\cdot }{\varphi },
\end{eqnarray}
where $H=\stackrel{\cdot }{a}/a$ is the Hubble scalar. The conservation
equation $\stackrel{\cdot }{\rho }_\varphi +3H(\rho _\varphi +p_\varphi )=0$
implies that the field equation for $\varphi $ is

\begin{equation}
\stackrel{\cdot \cdot }{\varphi }+\frac{dV}{d\varphi }-\frac 34H^2\varphi +3H%
\stackrel{\cdot }{\varphi }=0.
\end{equation}
The Friedman equation now reads

\begin{equation}
H^2=\frac{8\pi G}{3(1-\pi G\varphi ^2)}\left[ \frac 12\stackrel{\cdot }{%
\varphi }^2+V(\varphi ^2)\right] ,
\end{equation}
and the Raychaudhuri equation is
\begin{equation}
\stackrel{\cdot }{H}=-\frac{4\pi G}{1-\pi G\varphi ^2}\left( \stackrel{\cdot
}{\varphi }^2-\frac 12H\varphi \stackrel{\cdot }{\varphi }\right) .
\end{equation}

The field equations can be formulated as an autonomous system of two
differential equations by defining
\begin{eqnarray*}
u &=&\pi G\stackrel{\cdot }{\varphi }, \\
v &=&\sqrt{\pi G}\varphi , \\
h &=&\sqrt{\pi G}H(t),
\end{eqnarray*}
and
\[
f(v)=(\pi G)^2V(\varphi ^2),
\]
and introducing a new time coordinate
\[
\tau =\frac t{\sqrt{\pi G}}.
\]
The definition $v=\sqrt{\pi G}\varphi $ here is different from the one given
in [6], which leads to different results from [6]. Then, the field equations
can be rewritten as

\begin{eqnarray}
u_\tau &=&-\frac{df(v)}{dv}+\frac{u^2+2f(v)}{\left( 1-v^2\right) }v-2\sqrt{3}%
\sqrt{\frac{u^2+2f(v)}{1-v^2}}u, \\
v_\tau
&=&u,\;\;\;\;\;\;\;\;\;\;\;\;\;\;\;\;\;\;\;\;\;\;\;\;\;\;\;\;\;\;\;\;\;\;\;%
\;\;\;\;\;\;\;
\end{eqnarray}
\begin{equation}
h=\frac 2{\sqrt{3}}\sqrt{\frac{u^2+2f(v)}{1-v^2},}
\end{equation}
and
\begin{equation}
h_\tau =-\frac 4{\left( 1-v^2\right) }\left( u^2-\sqrt{\frac{u^2+2f(v)}{%
3\left( 1-v^2\right) }}uv\right) ,
\end{equation}
where $u_\tau =du/d\tau $. These equations (5,6) of $u$ and $v$ decouple
from $h$ and then define a dynamical flow in a large phase volume, the
behavior can be analyzed qualitatively by standard techniques from the
theory of ordinary differential equations. Of particular interest are the
exact solutions which define the fixed points of the system and their
stability around these fixed points. The fixed points of the system can be
obtained by solving $u_\tau =0$ and $v_\tau =0$ for $u$ and $v$:

\begin{eqnarray}
\frac{df(v)}{dv} &=&\frac{2vf(v)}{\left( 1-v^2\right) },  \nonumber \\
u &=&0,
\end{eqnarray}
which only corresponds to the equation (5.16) in [6]. The equation $\frac{%
df(v)}{dv}=2f(v)v/(1-v^2)$ can in principle have infinitely many solutions.
For example, $f(v)=c/\left( 1-v^2\right) $ solve this equation for all
values of $v$ and in that case we would encounter a fixed line.

The Jacobian of the system (5,6) is
\begin{equation}
{\cal M}=\left(
\begin{array}{ll}
\frac{\partial u_\tau }{\partial u} & \frac{\partial u_\tau }{\partial v} \\
\frac{\partial v_\tau }{\partial u} & \frac{\partial v_\tau }{\partial v}
\end{array}
\right)
\end{equation}
where
\begin{eqnarray}
\frac{\partial u_\tau }{\partial u} &=&\frac{2uv}{\left( 1-v^2\right) }-2%
\sqrt{3}\sqrt{\frac{u^2+2f(v)}{1-v^2}}-\frac{2\sqrt{3}}{\sqrt{\left(
1-v^2\right) \left( u^2+2f(v)\right) }}u^2,  \nonumber \\
\frac{\partial u_\tau }{\partial v} &=&-\frac{d^2f(v)}{dv^2}+\frac 2{\left(
1-v^2\right) }\frac{df(v)}{dv}v+\frac{u^2+2f(v)}{\left( 1-v^2\right) }+\frac{%
2u^2+4f(v)}{\left( 1-v^2\right) ^2}v^2  \nonumber \\
&&-2\sqrt{3}\frac 1{\sqrt{1-v^2}\sqrt{u^2+2f(v)}}\frac{df(v)}{dv}u-2\sqrt{3}%
\frac{\sqrt{u^2+2f(v)}}{\left( \sqrt{1-v^2}\right) ^3}uv,  \nonumber \\
\frac{\partial v_\tau }{\partial u} &=&1,\frac{\partial v_\tau }{\partial v}%
=0.
\end{eqnarray}
At the fixed point (9) we have
\begin{eqnarray*}
\frac{\partial u_\tau }{\partial u} &=&-2\sqrt{3}\sqrt{\frac{2f(v)}{1-v^2}},%
\frac{\partial u_\tau }{\partial v}=0, \\
\frac{\partial v_\tau }{\partial u} &=&1,\frac{\partial v_\tau }{\partial v}%
=0.
\end{eqnarray*}
The Jacobian ${\cal M}$ has the eigenvalues:$-2\sqrt{3}\sqrt{\frac{2f(v)}{%
1-v^2}},0$. For this nonhyperbolic fixed point the linearization theorem
does not yield any information about the stability of it and therefore, the
center manifold theorem is needed. The theorem shows that the qualitative
behavior in a neighborhood of a nonhyperbolic fixed point $p$ is determined
by its behavior on the center manifold near $p$. Since the dimension of the
center manifold is generally smaller than the dimension of the dynamical
system, this greatly simplifies the problem.

We shift the fixed point to ($0,0$) by setting
\[
v=\overline{v}+v_c,v_c=-2\sqrt{3}\sqrt{\frac{2f(v)}{1-v^2}}.
\]
and write the equation (5,6) as, up to second order,
\begin{eqnarray}
u_\tau &=&-\frac{2\sqrt{3}\sqrt{2f(v_c)}}{\sqrt{\left( 1-v_c^2\right) }}u+%
\frac{v_c}{\left( 1-v_c^2\right) }u^2-\frac{4\sqrt{6f(v_c)}v_c}{\sqrt{\left(
1-v_c^2\right) ^3}}u\overline{v}-12\frac{v_c\left( 1+v_c^2\right) f(v_c)}{%
\left( 1-v_c^2\right) ^3}\overline{v}^2,  \nonumber \\
\overline{v}_\tau
&=&u,\;\;\;\;\;\;\;\;\;\;\;\;\;\;\;\;\;\;\;\;\;\;\;\;\;\;\;\;\;\;\;\;\;\;\;%
\;\;\;\;\;\;\;\;\;\;\;\;\;\;\;\;\;\;\;
\end{eqnarray}
where
\[
\frac{df(v)}{dv}=\frac{2vf(v)}{\left( 1-v^2\right) },
\]
has been used.

The Jacobi decomposition
\begin{eqnarray*}
{\cal M} &=&\left(
\begin{array}{ll}
-2\sqrt{3}\sqrt{\frac{2f(v_c)}{1-v_c^2}} & 0 \\
1 & 0
\end{array}
\right) \\
&=&\left(
\begin{array}{cc}
0 & 1 \\
\frac 1{2\sqrt{3}\sqrt{\frac{2f(v_c)}{1-v_c^2}}} & -\frac 1{2\sqrt{3}\sqrt{%
\frac{2f(v_c)}{1-v_c^2}}}
\end{array}
\right) \left(
\begin{array}{cc}
0 & 0 \\
0 & -2\sqrt{3}\sqrt{\frac{2f(v_c)}{1-v_c^2}}
\end{array}
\right) \left(
\begin{array}{cc}
1 & 2\sqrt{3}\sqrt{\frac{2f(v_c)}{1-v_c^2}} \\
1 & 0
\end{array}
\right) \\
&=&{\cal SJS}^{-1},
\end{eqnarray*}
gives the transformation
\begin{eqnarray}
u &=&y,  \nonumber \\
\overline{v} &=&\frac 1{2\sqrt{3}\sqrt{\frac{2f(v_c)}{1-v_c^2}}}x-\frac 1{2%
\sqrt{3}\sqrt{\frac{2f(v_c)}{1-v_c^2}}}y,
\end{eqnarray}

\begin{eqnarray}
x &=&u+2\sqrt{3}\sqrt{\frac{2f(v_c)}{1-v_c^2}}\overline{v},  \nonumber \\
y &=&u,
\end{eqnarray}
and then
\begin{eqnarray}
x_\tau &=&u_\tau +2\sqrt{3}\sqrt{\frac{2f(v_c)}{1-v_c^2}}\overline{v}_\tau ,
\nonumber \\
y_\tau &=&u_\tau .
\end{eqnarray}
The system (12) can be written in diagonal form
\begin{eqnarray}
x_\tau &=&F\left( x,y\right) ,  \nonumber \\
y_\tau &=&\lambda x+G\left( x,y\right) ,
\end{eqnarray}
where
\begin{eqnarray}
F\left( x,y\right) &=&-\frac{v_c\left( 1+v_c^2\right) }{2\left(
1-v_c^2\right) ^2}x^2-\frac{v_c\left( 1-3v_c^2\right) }{2\left(
1-v_c^2\right) ^2}xy+\frac{v_c\left( 5-7v_c^2\right) }{2\left(
1-v_c^2\right) ^2}y^2,  \nonumber \\
G\left( x,y\right) &=&-\frac{v_c\left( 1+v_c^2\right) }{2\left(
1-v_c^2\right) ^2}x^2-\frac{v_c\left( 1-3v_c^2\right) }{2\left(
1-v_c^2\right) ^2}xy+\frac{v_c\left( 5-7v_c^2\right) }{2\left(
1-v_c^2\right) ^2}y^2,
\end{eqnarray}
$\left( x,y\right) $ $\in $ ${\Bbb R}$ $\times $ ${\Bbb R}$, $\lambda =-%
\frac{2\sqrt{3}\sqrt{2f(v_c)}}{\sqrt{\left( 1-v_c^2\right) }}$ is a negative
eigenvalues of the matrix ${\cal J}$ and $F$, $G$ vanish at the origin ($0,0$%
) and have vanishing derivatives at ($0,0$). The center manifold theorem
asserts that there exists a 1-dimensional invariant local center manifold $%
W^c(0)$ of (16) tangent to the center subspace (the $y$ $=0$ space) at $0$.
Moreover, $W^c(0)$ can be represented as

\begin{equation}
W^c(0)=\left\{ \left( x,y\right) \in {\Bbb R}\times {\Bbb R}%
:y=y(x),|x|<\delta \right\} ;y(0)=0,y^{\prime }(0)=0,
\end{equation}
for $\delta $ sufficiently small (see [9], p. 155) where $y^{\prime }(x)=%
\frac{dy}{dx}$. The function $y(x)$ that defines the local center manifold
satisfies

\begin{equation}
y^{\prime }(x)F\left( x,y(x)\right) -\lambda y(x)-G\left( x,y(x)\right) =0,
\end{equation}
for $|x|<\delta $; and the flow on the center manifold $W^c(0)$ is defined
by the equation
\begin{equation}
x_\tau =F\left( x,y(x)\right) ,
\end{equation}
for all $x$ with $|x|<\delta $.

According to Theorem 3.2.2 in [10], if the origin x = 0 of (20) is stable
(resp. unstable) then the origin of (16) is also stable (resp. unstable).
Therefore, we have to find the local center manifold, i.e., the problem
reduces to the computation of $y(x)$.

The condition (18) allows for an approximation of $y(x)$ by a Taylor series
at $x$ $=0$. Since $y(0)=0,y^{\prime }(0)=0$, it is obvious that $y(x)$
commences with quadratic terms. We substitute
\[
y(x)=y_2x^2+y_3x^3+y_4x^4+\cdots
\]
into (18) and set the coefficients of like powers of $x$ equal to zero to
find the unknowns
\begin{eqnarray*}
y_2 &=&-\frac A\lambda =-\frac{v_c\left( 1+v_c^2\right) }{4\sqrt{%
6f(v_c)\left( 1-v_c^2\right) ^3}}, \\
y_3 &=&\left( B-2A\right) \frac A{\lambda ^2}=-\frac{v_c^2\left(
1+6v_c^2+5v_c^4\right) }{96\left( 1-v_c^2\right) ^3f(v_c)}, \\
y_4 &=&\left( 2B-C\right) \allowbreak \frac{A^2}{\lambda ^3}+\left(
5AB-6A^2-B^2\right) \frac A{\lambda ^3}=\frac{v_c^3\left( 1+v_c^2\right)
\left( 5-22v_c^2-43v_c^4\right) \sqrt{\left( 1-v_c^2\right) }}{384\sqrt{3}%
f(v_c)\sqrt{2f(v_c)}\left( 1-v_c^2\right) ^5}, \\
&&\cdots
\end{eqnarray*}

Therefore, (20) yields

\[
x_\tau =-\frac{v_c\left( 1+v_c^2\right) }{2\left( 1-v_c^2\right) ^2}x^2+%
\frac{v_c^2\left( 1-3v_c^2\right) \left( 1+v_c^2\right) }{8\left(
1-v_c^2\right) ^3\sqrt{6f(v_c)\left( 1-v_c^2\right) }}x^3+\frac{v_c^3\left(
3-11v_c^4\right) \left( 1+v_c^2\right) }{96f(v_c)\left( 1-v_c^2\right) ^5}%
x^4+\cdots
\]
means that this fixed point is stable when $v_c=0$ or a {\em saddle-node},
otherwise. It is important to note that this stability is independent of the
potential function $f(v)=(\pi G)^2V(\varphi ^2)$. The equation (8) indicate
the whole fixed (point) line (9) corresponds to de Sitter solutions, however
only the point $u=0,v=0$ is stable.

\section{Cosmology of Lorentz Invariant NSS}

Now we turn to cosmology of the Lorentz invariant NSS spinor $\psi $
introduced in [6]. By defining
\begin{eqnarray*}
\widetilde{\psi } &=&a^{3/2}\psi , \\
\widetilde{\Phi } &=&\overline{\widetilde{\psi }}\widetilde{\psi },%
\widetilde{\Psi }=\stackrel{\cdot }{\overline{\widetilde{\psi }}}\stackrel{%
\cdot }{\widetilde{\psi }}
\end{eqnarray*}
and
\[
\Phi =a^{-3}\widetilde{\Phi },\Psi =a^{-3}\widetilde{\Psi }.
\]
the energy density $\rho _\psi $ and pressure $p_\psi $ can be written as

\begin{eqnarray}
\rho _\psi &=&\Psi +V(\Phi ),  \nonumber \\
p_\psi &=&V^{\prime }(\Phi )\Phi -V(\Phi ),
\end{eqnarray}
where $V^{\prime }(\Phi )=dV/d\Phi $. Then $\Phi $ and $\Psi $ can be
treated as the dynamical variables cosmological. The field equations of them
are

\begin{eqnarray}
\stackrel{\cdot \cdot }{\widetilde{\Phi }} &=&2\left[ \widetilde{\Psi }%
-V^{\prime }(\Phi )\widetilde{\Phi }\right] , \\
\stackrel{\cdot }{\widetilde{\Psi }} &=&-V^{\prime }(\Phi )\stackrel{\cdot }{%
\widetilde{\Phi }},
\end{eqnarray}
The Friedman equation and the Raychaudhuri equation now read, respectively,

\begin{equation}
H^2=\frac \kappa 3[\Psi +V(\Phi )],
\end{equation}
and
\begin{equation}
3H^2+2\stackrel{\cdot }{H}=\kappa \left[ V(\Phi )-V^{\prime }(\Phi )\Phi
\right] .
\end{equation}

In contrast to [6], we do not search for the de Sitter solutions directly
using the condition $p_\psi =-\rho _\psi $. Instead, we will discuss the
exact solutions of the equations (22) and (23) given by fixed points of a
dynamical system and then obtain de Sitter solutions from (24). Using
\[
H=\frac{\stackrel{\cdot }{a}}a,\widetilde{\Phi }=a^3\Phi ,\stackrel{\cdot
\cdot }{\widetilde{\Psi }=a^3\Psi ,}
\]
and (24), (25) the equations (22) and (23) can be rewritten as
\begin{eqnarray}
\stackrel{\cdot \cdot }{\Phi } &=&-6\stackrel{\cdot }{\Phi }\sqrt{\frac %
\kappa 3[\Psi +V(\Phi )]}-3\kappa V(\Phi )\Phi -\frac 32\kappa \Psi \Phi
+\left( \frac 32\kappa \Phi -2\right) V^{\prime }(\Phi )\Phi -\left( \frac 32%
\kappa \Phi -2\right) \Psi ,  \nonumber \\
\stackrel{\cdot }{\Psi } &=&-3\Psi \sqrt{\frac \kappa 3[\Psi +V(\Phi )]}-%
\stackrel{\cdot }{\Phi }V^{\prime }(\Phi )-3\Phi V^{\prime }(\Phi )\sqrt{%
\frac \kappa 3[\Psi +V(\Phi )]},
\end{eqnarray}
which have decoupled from $H$. (25) becomes
\begin{equation}
\stackrel{\cdot }{H}=-\frac \kappa 2\left[ \Psi +V^{\prime }(\Phi )\Phi
\right] .
\end{equation}

Let
\[
\Theta =\stackrel{\cdot }{\Phi },
\]
then we obtain a dynamical system
\begin{eqnarray}
\stackrel{\cdot }{\Theta } &=&-2\sqrt{3\kappa }\Theta \sqrt{\Psi +V(\Phi )}%
-3\kappa V(\Phi )\Phi +\left( \frac 32\kappa \Phi -2\right) V^{\prime }(\Phi
)\Phi -\left( \frac 32\kappa \Phi -2\right) \Psi ,  \nonumber \\
\stackrel{\cdot }{\Phi } &=&\Theta  \nonumber \\
\stackrel{\cdot }{\Psi } &=&-\Theta V^{\prime }(\Phi )-\sqrt{3\kappa }\Phi
V^{\prime }(\Phi )\sqrt{\Psi +V(\Phi )}-\sqrt{3\kappa }\Psi \sqrt{\Psi
+V(\Phi )},
\end{eqnarray}
with fixed points

1) A:
\begin{eqnarray*}
-\left( \frac 32\kappa \Phi +2\right) V(\Phi )+\left( \frac 32\kappa \Phi
-2\right) V^{\prime }(\Phi )\Phi &=&0, \\
\Theta &=&0, \\
\Psi &=&-V(\Phi ),
\end{eqnarray*}

2) B:
\begin{eqnarray*}
-3\kappa V(\Phi )+3\kappa V^{\prime }(\Phi )\Phi -4V^{\prime }(\Phi ) &=&0,
\\
\Theta &=&0, \\
\Psi &=&-\Phi V^{\prime }(\Phi ),
\end{eqnarray*}

3) C:
\[
\Phi =0,\Theta =0,\Psi =0.
\]
The equation (27) indicates that the points B and C correspond to the de
Sitter solution
\[
\stackrel{\cdot }{H}=0.
\]

The stability of these fixed points is described by the Jacobian
\begin{equation}
{\cal M}=\left(
\begin{array}{lll}
\frac{\partial \stackrel{\cdot }{\Theta }}{\partial \Theta } & \frac{%
\partial \stackrel{\cdot }{\Theta }}{\partial \Phi } & \frac{\partial
\stackrel{\cdot }{\Theta }}{\partial \Psi } \\
\frac{\partial \stackrel{\cdot }{\Phi }}{\partial \Theta } & \frac{\partial
\stackrel{\cdot }{\Phi }}{\partial \Phi } & \frac{\partial \stackrel{\cdot }{%
\Phi }}{\partial \Psi } \\
\frac{\partial \stackrel{\cdot }{\Psi }}{\partial \Theta } & \frac{\partial
\stackrel{\cdot }{\Psi }}{\partial \Phi } & \frac{\partial \stackrel{\cdot }{%
\Psi }}{\partial \Psi }
\end{array}
\right) ,
\end{equation}
where
\begin{eqnarray*}
\frac{\partial \stackrel{\cdot }{\Theta }}{\partial \Theta } &=&-2\sqrt{%
3\kappa }\sqrt{\Psi +V(\Phi )}, \\
\frac{\partial \stackrel{\cdot }{\Theta }}{\partial \Phi } &=&-\frac{\sqrt{%
3\kappa }\Theta }{\sqrt{\Psi +V(\Phi )}}V^{\prime }(\Phi )-3\kappa V(\Phi
)-2V^{\prime }(\Phi )+\left( \frac 32\kappa \Phi -2\right) V^{\prime \prime
}(\Phi )\Phi -\frac 32\kappa \Psi , \\
\frac{\partial \stackrel{\cdot }{\Theta }}{\partial \Psi } &=&-\frac{\sqrt{%
3\kappa }\Theta }{\sqrt{\Psi +V(\Phi )}}+\left( 2-\frac 32\kappa \Phi
\right) ,
\end{eqnarray*}
\[
\frac{\partial \stackrel{\cdot }{\Phi }}{\partial \Theta }=1,\frac{\partial
\stackrel{\cdot }{\Phi }}{\partial \Phi }=0,\frac{\partial \stackrel{\cdot }{%
\Psi }}{\partial \Psi }=0,
\]
\begin{eqnarray*}
\frac{\partial \stackrel{\cdot }{\Psi }}{\partial \Theta } &=&-V^{\prime
}(\Phi ), \\
\frac{\partial \stackrel{\cdot }{\Psi }}{\partial \Phi } &=&-\Theta
V^{\prime \prime }(\Phi )-\sqrt{3\kappa }\left( \Phi V^{\prime \prime }(\Phi
)+V^{\prime }(\Phi )\right) \sqrt{\Psi +V(\Phi )}-\frac{\sqrt{3\kappa }}2%
\frac{\Phi \left( V^{\prime }(\Phi )\right) ^2+\Psi V^{\prime }(\Phi )}{%
\sqrt{\Psi +V(\Phi )}}, \\
\frac{\partial \stackrel{\cdot }{\Psi }}{\partial \Psi } &=&-\sqrt{3\kappa }%
\sqrt{\Psi +V(\Phi )}-\frac{\sqrt{3\kappa }}2\frac{\Phi V^{\prime }(\Phi
)+\Psi }{\sqrt{\Psi +V(\Phi )}}.
\end{eqnarray*}

At B:
\begin{eqnarray*}
-3\kappa V(\Phi _c)+3\kappa V^{\prime }(\Phi _c)\Phi _c-4V^{\prime }(\Phi
_c) &=&0, \\
\Theta _{,} &=&0, \\
\Psi _c &=&-\Phi _cV^{\prime }(\Phi _c),
\end{eqnarray*}
the Jacobian of ${\cal M}$ has the eigenvalues: $-4\sqrt{-V^{\prime }(\Phi
_c)}$, $-\sqrt{-V^{\prime }(\Phi _c)}-\sqrt{-V^{\prime }(\Phi _c)-V^{\prime
\prime }(\Phi _c)\Phi _c\left( 2-\frac 32\kappa \Phi _c\right) }$, $-\sqrt{%
-V^{\prime }(\Phi _c)}+\sqrt{-V^{\prime }(\Phi _c)-V^{\prime \prime }(\Phi
_c)\Phi _c\left( 2-\frac 32\kappa \Phi _c\right) }$.

i) In the case
\begin{equation}
V^{\prime }(\Phi _c)<0,
\end{equation}
when
\begin{equation}
0<\Phi _c<\frac 4{3\kappa },V^{\prime \prime }(\Phi _c)>0,
\end{equation}
or
\begin{equation}
\Phi _c>\frac 4{3\kappa },V^{\prime \prime }(\Phi _c)<0,
\end{equation}
the point B is stable, otherwise or $\Phi _c<0$, it is unstable. However, if
\[
V^{\prime \prime }(\Phi _c)=0,
\]
or
\[
\Phi _c=0,
\]
the eigenvalues are $-4\sqrt{-V^{\prime }(\Phi _c)}$, $-2\sqrt{-V^{\prime
}(\Phi _c)}$, $0$, and then B is a saddle-node i.e. it behaves like a saddle
or an attractor depending on the direction from which the orbit approaches.

ii) In the case
\[
V^{\prime }(\Phi _c)=0,
\]
the eigenvalues are $0$, $-\sqrt{-V^{\prime \prime }(\Phi _c)\Phi _c\left( 2-%
\frac 32\kappa \Phi _c\right) }$, $\sqrt{-V^{\prime \prime }(\Phi _c)\Phi
_c\left( 2-\frac 32\kappa \Phi _c\right) }$.

When
\[
\Phi _c>\frac 4{3\kappa },V^{\prime \prime }(\Phi _c)>0,-V^{\prime \prime
}(\Phi _c)\Phi _c\left( 2-\frac 32\kappa \Phi _c\right) >0,
\]
or
\[
\frac 4{3\kappa }>\Phi _c>0,V^{\prime \prime }(\Phi _c)<0,-V^{\prime \prime
}(\Phi _c)\Phi _c\left( 2-\frac 32\kappa \Phi _c\right) >0,
\]
the point B is unstable, while
\[
\Phi _c>\frac 4{3\kappa },V^{\prime \prime }(\Phi _c)<0,-V^{\prime \prime
}(\Phi _c)\Phi _c\left( 2-\frac 32\kappa \Phi _c\right) <0,
\]
or
\[
\frac 4{3\kappa }>\Phi _c>0,V^{\prime \prime }(\Phi _c)>0,-V^{\prime \prime
}(\Phi _c)\Phi _c\left( 2-\frac 32\kappa \Phi _c\right) <0,
\]
it is a center.

When
\[
\Phi _c<0,V^{\prime \prime }(\Phi _c)>0,
\]
B is unstable, while
\[
\Phi _c<0,V^{\prime \prime }(\Phi _c)<0,
\]
it is a center.

When
\[
\Phi _c=0,
\]
the point B is a center.

iii) In the case
\[
V^{\prime }(\Phi _c)>0,
\]
when
\[
-V^{\prime }(\Phi _c)>V^{\prime \prime }(\Phi _c)\Phi _c\left( 2-\frac 32%
\kappa \Phi _c\right) ,
\]
B is unstable, while
\[
-V^{\prime }(\Phi _c)\leq V^{\prime \prime }(\Phi _c)\Phi _c\left( 2-\frac 32%
\kappa \Phi _c\right) ,
\]
it is a center. These results indicate that the stability of the fixed point
B is dependent on $\Phi _c$, $V^{\prime }(\Phi _c)$ and $V^{\prime \prime
}(\Phi _c)$ but independent of $V(\Phi _c)$.

At C:
\[
\Phi _c=0,\Theta _c=0,\Psi _c=0,
\]
the Jacobian has the eigenvalues: $-\sqrt{3\kappa V(0)},-\sqrt{3\kappa V(0)}%
-2\sqrt{-V^{\prime }(0)}\allowbreak ,-\sqrt{3\kappa V(0)}+2\sqrt{-V^{\prime
}(0)}\allowbreak $.

i) In the case
\begin{equation}
V(0)>0,
\end{equation}
when

\begin{equation}
V^{\prime }(0)\geq 0,
\end{equation}
C is stable. When

\begin{equation}
V^{\prime }(0)<0,-\sqrt{3\kappa V(0)}+2\sqrt{-V^{\prime }(0)}<0,
\end{equation}
C is stable, while
\[
V^{\prime }(0)<0,-\sqrt{3\kappa V(0)}+2\sqrt{-V^{\prime }(0)}>0,
\]
it is unstable. When
\[
V^{\prime }(0)<0,-\sqrt{3\kappa V(0)}+2\sqrt{-V^{\prime }(0)}=0,
\]
the Jacobin has a zero eigenvalue: $-\sqrt{3\kappa V(0)},-2\sqrt{3\kappa V(0)%
}\allowbreak ,0$, and then$\;$C is a saddle-node.

ii) In the case
\[
V(0)<0,
\]
when
\[
V^{\prime }(0)\geq 0,
\]
C is a center, while
\[
V^{\prime }(0)<0,
\]
it is unstable.

iii) In the case
\[
V(0)=0,
\]
when
\[
V^{\prime }(0)\geq 0,
\]
C is a center, while
\[
V^{\prime }(0)<0,
\]
C is unstable. The stability of the fixed point C depends on $V(0)$ and $%
V^{\prime }(0)$.

\section{Conclusions}

Cosmology of ELKO and Lorentz Invariant NSS has been investigated using
dynamical system method. The exact solutions described by fixed points of
the dynamical system and their stability have been discussed. Some stable
solutions corresponding to de Sitter universe have been obtained. In the
cosmology of ELKO there exists a stable fixed point corresponding to a de
Sitter universe while in the cosmology of Lorentz Invariant NSS we encounter
some stable fixed lines corresponding to de Sitter solutions. These results
and conclusions, especially the de Sitter solutions and their behavior are
different from the ones in [6].

\begin{acknowledgments}
This work was supported by the National Natural Science Foundation
of China (Grant Nos.~10705041 and 10975032).
\end{acknowledgments}

\vskip1cm

[1] D. V. Ahluwalia-Khalilova and D. Grumiller, Phys. Rev. {\bf
D72,} 067701 (2005), [hep-th/0410192]; JCAP {\bf 0507} 012,
[hep-th/0412080].

[2] C. G. Boehmer, Annalen Phys. (Leipzig) {\bf 16}, 38 (2007),
[gr-qc/0607088]; Annalen Phys. (Leipzig) {\bf 16, } 325, (2007)
[gr-qc/0701087]; L. Fabbri, Mod. Phys. Lett. {\bf A25,} 151 (2010),
[arXiv:0911.2622].

[3] C. G. Boehmer and D. F. Mota, Phys. Lett. {\bf B663,} 168 (2008),
[arXiv:0710.2003]; C. G. Boehmer and J. Burnett, Phys. Rev. {\bf D78},
104001 (2008), [arXiv:0809.0469]; C. G. Boehmer, Phys. Rev. {\bf D77},
123535 (2008), [arXiv:0804.0616]; C. G. Boehmer and J. Burnett, Mod. Phys.
Lett. {\bf A25, }101 (2010), [arXiv:0906.1351]; C. G. Boehmer and J.
Burnett, arXiv:1001.1141, {\it Dark Spinors}.

[4] R. da Rocha and W. A. Rodrigues, Jr., Mod. Phys. Lett. {\bf A21,} 65
(2006), [math-ph/0506075]; R. da Rocha and J. M. Hoff da Silva,
arXiv:0811.2717, {\it ELKO flagpole and flag-dipole spinor fields, and the
instanton Hopf fibration}; R. da Rocha and J. M. Hoff da Silva, Int. J.
Geom. Meth. Mod. Phys. {\bf 6,} 461 (2009), [arXiv:0901.0883]; J. M. Hoff da
Silva and R. da Rocha, Int. J. Mod. Phys. {\bf A24,} 3227 (2009),
[arXiv:0903.2815].

[5] D. Gredat and S. Shankaranarayanan, JCAP 1001, 008 (2010),
[arXiv:0807.3336]; S. Shankaranarayanan,, arXiv:0905.2573, {\it What-if
inflaton is a spinor condensate?}; S. Shankaranarayanan, arXiv:1002.1128,
{\it Dark spinor driven inflation};

[6] C. G. Boehmer, J. Burnett, D. F. Mota, and D. J. Shaw, arXiv:1003.3858,
{\it Dark spinor models in gravitation and cosmology.}

[7] J. Wainwright and G. F. R. Ellis, {\it Dynamical System in Cosmology,}
(Cambridge University Press, Cambridge, 1997); A. A. Coley,
arXiv:gr-qc/9910074, Dynamical Systems in Cosmology.

[8] C. G. Boehmer, G. Caldera-Cabral, R. Lazkoz, and R. Maartens, Phys.Rev.
{\bf D78}, 023505 (2008), arXiv: 0801.1565.

[9] L. Perko, Differential Equations and Dynamical Systems,
(Springer-Verlag, 1991).

[10] J. Guckenheimer and P. Holmes, Nonlinear Oscillations, Dynamical
Systems and Bifurcations of Vector Fields, (Springer-Verlag, New York, 1983).


\end{document}